\documentclass[DIV=12,numbers=noenddot]{scrartcl}
\usepackage[utf8]{inputenc}
\usepackage{slashed}
\usepackage{booktabs}
\usepackage{multirow}
\usepackage{graphicx}
\usepackage[section]{placeins}
\usepackage{longtable}
\usepackage{tabu}
\usepackage[usenames]{xcolor}
\usepackage{lmodern}
\usepackage{amsmath,amssymb}
\usepackage[usenames]{xcolor}
\usepackage{scalefnt}
\usepackage{xspace}
\usepackage{listings}
\usepackage{booktabs}

\usepackage{authblk}
\usepackage[font=small,labelfont=bf,format=plain,margin=0.05\textwidth]{caption}
\usepackage{subcaption}

\renewcommand{\imath}{\mathrm{i}}
\newcommand{\muu}[1]{\mu_u^{\text{eff,}#1}}
\newcommand{\mud}[1]{\mu_d^{\text{eff,}#1}}

\usepackage[
  pdftitle={Interpreting 95 GeV di-photon/bbbar excesses as a lightest Higgs boson of the MRSSM},
  pdfauthor={Jan Kalinowski, Wojciech Kotlarski},
  pdfkeywords={FlexibleSUSY,Higgs boson,95 GeV excess},
  bookmarks=true, linktocpage, colorlinks=true, allbordercolors=white,
  allcolors=blue]{hyperref}
\usepackage{orcidlink} % must come AFTER hyperref

\usepackage[
  backend=biber,
  bibencoding=utf8,
  sorting=none,
  url=true,
  doi=false,
  style=phys,
  biblabel=brackets,
  subentry=false,
  eprint=true,
  maxbibnames=5
]{biblatex}
\AtEveryBibitem{%
  \clearfield{note}%
}

\title{\boldmath Interpreting 95 GeV di-photon/$b\bar{b}$ excesses as a lightest Higgs boson of the MRSSM}
\date{}
\author[1]{Jan Kalinowski\orcidlink{0000-0001-5618-0141}\thanks{\href{mailto:jan.kalinowski@fuw.edu.pl}{jan.kalinowski@fuw.edu.pl}}}
\author[2]{Wojciech Kotlarski\orcidlink{0000-0002-1191-6343}\thanks{\href{mailto:wojciech.kotlarski@ncbj.gov.pl}{wojciech.kotlarski@ncbj.gov.pl}}}

\affil[1]{Faculty of Physics, University of Warsaw, Pasteura 5, 02-093 Warsaw, Poland}
\affil[2]{National Centre for Nuclear Research, Pasteura 7, 02-093 Warsaw, Poland}

\newcommand{\SciPy}{\texttt{Sci\-Py}\@\xspace}
\newcommand{\FS}{\texttt{Flex\-ib\-le\-SUSY}\@\xspace}
\newcommand{\PySLHA}{\texttt{PySLHA}\@\xspace}
\newcommand{\SLHA}{\texttt{SLHA}\@\xspace}
\newcommand{\MO}{\texttt{micrOMEGAs}\@\xspace}
\newcommand{\SA}{\texttt{SARAH}\@\xspace}
\newcommand{\HB}{\texttt{Higgs\-Bounds}\@\xspace}
\newcommand{\HT}{\texttt{Higgs\-Tools}\@\xspace}
\newcommand{\HS}{\texttt{Higgs\-Signals}\@\xspace}
\newcommand{\LL}{\texttt{Lilith}\@\xspace}

\newcommand{\muzbb}{\mu_{Z b\bar{b}}}
\newcommand{\muaa}{\mu_{\gamma \gamma}}

\begin{document}
\maketitle
\begin{abstract}
The Minimal R-symmetric Supersymmetric Standard Model (MRSSM) is a well motivated BSM model which can accommodate the observed 125 GeV Higgs boson in agreement with electroweak precision observables, in particular with the $W$ boson mass and $T$ parameter.
In the 2016 paper we showed that the SM-like 125 GeV Higgs state can be also realised as the second-to-lightest scalar of the MRSSM, leaving room for another sub-100 GeV state.
Motivated by the recent ATLAS and CMS observation of the di-photon excess at a mass of around 95 GeV we investigate the possibility whether this could be the lightest CP-even MRSSM scalar in a variation
of our benchmarks presented in the 2016 work.
We show that such  a state can also simultaneously explain the excess in the $b\bar{b}$ final state observed around the same mass value at LEP.
Due to the R-symmetric nature of the model, a light singlet-like Higgs state leads necessarily to a light bino-singlino Dirac dark matter candidate, which can give a correct relic density while evading current experimental bounds.
Dark matter and LHC searches place further bounds on this scenario and point to parameter regions which are viable and of interest for the LHC Run III and upcoming dark matter experiments.

\end{abstract}

\tableofcontents

\section{Introduction}

After the discovery of a Standard Model (SM)-like Higgs boson at the LHC in 2012, it remains an open question whether this is the only fundamental scalar particle or whether it  actually heralds a beyond the SM theory (BSM) with an extended scalar sector.
Such sectors are  notorious in many BSM theories that aim to address  at least some unsolved puzzles of the SM, like the hierarchy of scales, neutrino masses, dark matter, baryogenesis etc.

In the absence of any other direct experimental BSM signal (new coloured particles, heavy fermions etc.), a careful study of the Higgs sector properties might be the only way for inferring properties of new physics in coming years. Any BSM scenario must allow to accommodate  a SM-like 125 GeV state ($h^\text{SM}_{125}$) with properties consistent with current  measurements. While the couplings of the $h^\text{SM}_{125}$ state to gauge bosons and third generation fermions and muons have been checked to be SM-like within experimental errors, there is still some room for small deviations. And the $h^\text{SM}_{125}$ self-interactions are still unconstrained.

However, an intriguing question is whether there are additional scalar particles, possibly even with smaller mass.
Searches for scalar particles below 125 GeV have been performed in the past  at LEP and Tevatron and now at the LHC. Results based on Run 1 and the first years of Run 2 data collected by the CMS in di-photon channel showed an excess of 2.9$\sigma$ (local) for a hypothetical mass of 95.3 GeV \cite{CMS-PAS-HIG-17-013}.
Recently CMS confirmed the excess of di-photon events at the 2.9$\sigma$ level for a mass of 95.4 GeV based on full Run 2 data set  \cite{CMS-PAS-HIG-20-002}. ATLAS search found 1.7$\sigma$ (local) around 95 GeV \cite{ATLAS-CONF-2023-035}. It is worth recalling an excess around 95 GeV in the $b\bar b$ channel in LEP searches \cite{LEPWorkingGroupforHiggsbosonsearches:2003ing}.
An analogous small excess has also been seen in the $\tau \bar{\tau}$ channel in CMS \cite{CMS:2022goy}.

These findings have triggered speculations that at least some of those excesses could arise from the  production of a new particle (see for example \cite{Biekotter:2023jld,Cao:2023gkc} and references therein).

In the 2016 paper \cite{Diessner:2015iln} we considered a possibility of a light (below 125 GeV) singlet-like scalar in the Minimal R-symmetric Supersymmetric model (MRSSM) \cite{Kribs:2007ac}.
The MRSSM is an attractive alternative to the MSSM, elegantly solving some  of its shortcomings while also significantly altering its phenomenology.
For example, the SUSY flavour problem is alleviated as R-symmetry forbids left-right squark mixing which is responsible for many flavour-violating interactions.
Scalar quarks can be lighter than in the MSSM as the collider limits for squark masses are weakened \cite{Heikinheimo:2011fk, Kribs:2012gx, Kribs:2013oda, Diessner:2017ske, Diessner:2019bwv,Borschensky:2024zdg} when simultaneously the electroweak and Higgs sectors contain new interactions which can push the SM-like Higgs boson mass up to the observed value for smaller top-squark masses \cite{Bertuzzo:2014bwa, Diessner:2014ksa, Diessner:2015yna, Diessner:2015iln}.
Those interactions can also contribute to the W-boson mass~\cite{Diessner:2014ksa, Diessner:2015iln, Athron:2022isz}.
The MRSSM also contains various possibilities to explain dark matter~\cite{Belanger:2009wf, Chun:2009zx, Buckley:2013sca} as well as predicts new color-octet scalars \cite{Choi:2008ub,Plehn:2008ae,Goncalves-Netto:2012gvn,Kotlarski:2016zhv,Darme:2018dvz} and Dirac gauginos \cite{Choi:2010gc,Choi:2009ue,Chalons:2018gez}.
The lepton flavour properties have also been analysed in Refs.~\cite{Dudas:2013gga, Fok:2010vk}, and in relation to the muon $\mathit{g}\!-\!2$ in Ref.~\cite{Kotlarski:2019muo}.

However, it remained an open question whether the aforementioned experimental deviations could be accommodated within the MRSSM.
In this work we will show that this is indeed the case.

The paper is structured as follows.
In Sec.~\ref{sec:hints} we recap experimental hints pointing to the existence of a $\sim$95 GeV scalar resonance.
In Sec.~\ref{sec:mrssm} we review the basics of the MRSSM and describe the model setup explaining the LHC and LEP excesses --- we also propose two benchmark points.
In Sec.~\ref{sec:experimental_constraints} we confront the benchmark points of Sec.~\ref{sec:mrssm} with Higgs physics, dark matter and general LHC constraints, before concluding in Sec.~\ref{sec:conclussions}.

\section{Hints of a new scalar with  the mass between 95 and 100 GeV}
\label{sec:hints}

In this section we briefly summarize the experimental hints for a $\sim95$ GeV scalar resonance.
First, there is a long standing LEP anomaly observed in the $Z b \bar{b}$ final state in the $95 \lesssim m_{b\bar{b}} \lesssim 100$ invariant mass window \cite{LEPWorkingGroupforHiggsbosonsearches:2003ing}.
This anomaly can be explained for example by a scalar state $s$ with a mass of 98 GeV whose combined production and branching ratio is roughly an order of magnitude smaller than that of a hypothetical SM-like Higgs $h_{98}^\text{SM}$ of the same mass \cite{Cao:2016uwt}
\begin{equation}
\mu_{Zb\bar{b}} = \frac{\sigma\left(e^+ e^- \to Z^* s \to Z b\bar{b} \right)}{\sigma\left(e^+ e^- \to Z^* h^\text{SM}_{98}\to Z b \bar{b} \right)}  = 0.117 \pm 0.057.
\end{equation}
Hints of such a state emerged also in the 8 and 13 TeV LHC data \cite{CMS-PAS-HIG-17-013,CMS-PAS-HIG-20-002,ATLAS-CONF-2023-035}.
A naive combination of ATLAS and CMS results points to the $3.1\sigma$ (local) excess with mass 95.4 GeV and signal strength \cite{Biekotter:2023oen}
\begin{align}
 \mu_{\gamma \gamma}^\text{ATLAS+CMS} = \frac{\sigma(gg\to s \to \gamma \gamma)}{\sigma (gg \to h^\text{SM}_{95.4} \to \gamma \gamma)} = 0.24^{+0.09}_{-0.08}.
\end{align}
Finally, it is worthwhile mentioning that there is also a  hint of an analogous excess in the $\tau\bar{\tau}$ channel with $\mu_{\tau \tau} = 1.2 \pm 0.5$ \cite{CMS:2022goy}.

As the LEP excess is very broad, both $\mu_{Zb\bar b}$ and $\mu_{\gamma \gamma}$ can be addressed simultaneously by a single state.
Such scenario is possible in many BSM models with an extended Higgs sector.
In non-supersymmetric models it has been analysed in a plethora of models (see for example \cite{Chen:2023bqr,Biekotter:2023oen,Azevedo:2023zkg}).
In SUSY however, it cannot be realized in its minimal version \cite{Bechtle:2016kui} but 
 can be  accommodated in extended SUSY, like the NMSSM (e.g. \cite{Cao:2016uwt,Cao:2023gkc}), $\mu\nu$SSM (e.g.  \cite{Biekotter:2017xmf,Liu:2024cbr}) or, as we will show, the MRSSM.

\section{A light CP-even scalar in the MRSSM}
\label{sec:mrssm}

\begin{table}[t]
\begin{center}
\begin{tabular}{c|l|l||l|l|l|l}
%\hline
\multicolumn{1}{c}{Field} & \multicolumn{2}{c}{Superfield} &
                              \multicolumn{2}{c}{Boson} &
                              \multicolumn{2}{c}{Fermion} \\
\hline 
 \phantom{\rule{0cm}{5mm}}Gauge Vector    &\, $\hat{g},\hat{W},\hat{B}$        \,& \, $\;\,$ 0 \,
          &\, $g,W,B$                 \,& \, $\;\,$ 0 \,
          &\, $\tilde{g},\tilde{W}\tilde{B}$             \,& \, +1 \,  \\
Matter   &\, $\hat{l}, \hat{e}$                    \,& \,\;+1 \,
          &\, $\tilde{l},\tilde{e}^*_R$                 \,& \, +1 \,
          &\, $l,e^*_R$                                 \,& $\;\;\,$\,\;0 \,    \\
          &\, $\hat{q},{\hat{d}},{\hat{u}}$       \,& \,\;+1 \,
          &\, $\tilde{q},{\tilde{d}}^*_R,{\tilde{u}}^*_R$ \,& \, +1 \,
          &\, $q,d^*_R,u^*_R$                             \,& $\;\;\,$\,\;0 \,    \\
 $H$-Higgs    &\, ${\hat{H}}_{d,u}$   \,& $\;\;\,$\, 0 \,
          &\, $H_{d,u}$               \,& $\;\;\,$\, 0 \,
          &\, ${\tilde{H}}_{d,u}$     \,& \, $-$1 \, \\ \hline
\phantom{\rule{0cm}{5mm}} R-Higgs    &\, ${\hat{R}}_{d,u}$   \,& \, +2 \,
          &\, $R_{d,u}$               \,& \, +2 \,
          &\, ${\tilde{R}}_{d,u}$     \,& \, +1 \, \\
  Adjoint Chiral  &\, $\hat{\cal O},\hat{T},\hat{S}$     \,& \, $\;\,$ 0 \,
          &\, $O,T,S$                \,& \, $\;\,$ 0 \,
          &\, $\tilde{O},\tilde{T},\tilde{S}$          \,& \, $-$1 \,  \\
%\hline
\end{tabular}
\end{center}
\caption{The R-charges of the superfields and the corresponding bosonic and
             fermionic components.
        }
\label{tab:Rcharges}
\end{table}

MRSSM is very well motivated, with phenomenology distinctly different from the MSSM as it contains a continuous,  unbroken at the low scale $U_R(1)$ R-symmetry under which both superfields and Grassmannian variables $\theta$ are charged ~\cite{Salam:1974xa,Fayet:1974pd}.
At the level of the Lagrangian, the $U_R(1)$ transformation of the coordinate $\theta\to e^{i\varphi}\theta$ implies that the superpotential $W$ has R-charge +2.
Assigning zero R-charges to the SM components of supermultiplets (in analogy to R-parity) forbids then Majorana gaugino and higgsino mass terms.
Therefore an enlarged field content is needed to account for non-vanishing of those masses.

In  the MRSSM  the standard MSSM matter, Higgs and gauge superfields are augmented by the R-charge 0 adjoint chiral
superfields $\hat{\cal O},\hat{T},\hat{S}$  for each gauge $SU(3)_c$, $SU(2)_L$, $U_Y(1)$ sector, respectively, and two Higgs iso-doublet superfields
${\hat{R}}_{d,u}$ with R-charge 2 (see Tab.~\ref{tab:Rcharges} for summary of the MRSSM particle content including the R-charge assignements).
The MRSSM superpotential takes the form of
\begin{align}
\nonumber W = & \mu_d\,\hat{R}_d \cdot \hat{H}_d\,+\mu_u\,\hat{R}_u\cdot\hat{H}_u\,+\Lambda_d\,\hat{R}_d\cdot \hat{T}\,\hat{H}_d\,+\Lambda_u\,\hat{R}_u\cdot\hat{T}\,\hat{H}_u\,\\ 
 & +\lambda_d\,\hat{S}\,\hat{R}_d\cdot\hat{H}_d\,+\lambda_u\,\hat{S}\,\hat{R}_u\cdot\hat{H}_u\,
 - Y_d \,\hat{d}\,\hat{q}\cdot\hat{H}_d\,- Y_e \,\hat{e}\,\hat{l}\cdot\hat{H}_d\, +Y_u\,\hat{u}\,\hat{q}\cdot\hat{H}_u\, ,
\label{eq:superpot}
 \end{align} 
where standard notation is used for the MSSM-like fields.
This is then supplemented by a $D$-term SUSY breaking Lagrangian
\begin{align}
 V_D= \, & M_B^D (\tilde{B}\,\tilde{S}-\sqrt{2} \mathcal{D}_B\, S)+
M_W^D(\tilde{W}^a\tilde{T}^a-\sqrt{2}\mathcal{D}_W^a T^a)+
M_O^D(\tilde{g}^a\tilde{O}^a-\sqrt{2}\mathcal{D}_g^a O^a)
 %-\sqrt{2} (M_B^D \mathcal{D}_B S + M_W^D \mathcal{D}_W^a T^a+ M_g^D \mathcal{D}_B^a O^a)
+ \mbox{h.c.}\,,
\label{eq:potdirac}
\end{align}
which generates the Dirac mass terms for gauginos.

Superfields $\hat H_{u,d}$, $S$ and $T$ present in Eq.~\ref{eq:superpot} mix to form physical Higgs boson.
The mass matrix of the CP-even neutral Higgs bosons in the weak basis $(\phi_d,\phi_u,\phi_S,\phi_T)$  is then given by~\cite{Diessner:2014ksa}
\begin{equation}\label{eq:scalarmassmatrix}
\mathcal{M}_{H^0}=
\begin{pmatrix}
\mathcal{M}_{\text{MSSM}}
&\mathcal{M}_{21}^T\\
\mathcal{M}_{21} &
\mathcal{M}_{22}
\\
\end{pmatrix}
\end{equation}
with the  sub-matrices 
\begin{align}
\mathcal{M}_{\text{MSSM}}&= 
\begin{pmatrix}
 m_Z^2 c_\beta^2+m_A^2 s_\beta^2 \; & -(m_Z^2 + m_A^2)s_\beta c_\beta  \\
  -(m_Z^2 + m_A^2)s_\beta c_\beta \; &  m_Z^2 s_\beta^2+m_A^2 c_\beta^2\\
\end{pmatrix}
\;, \notag\\
\mathcal{M}_{22}&= 
\begin{pmatrix} 4 (M_B^D)^2+m_S^2+\frac{\lambda_d^2 v_d^2+\lambda_u^2 v_u^2}{2} \;
& \frac{\lambda_d \Lambda_d v_d^2-\lambda_u \Lambda_u v_u^2}{2 \sqrt{2}} \\
 \frac{\lambda_d \Lambda_d v_d^2-\lambda_u \Lambda_u v_u^2}{2 \sqrt{2}} \;
 & 4 (M_W^D)^2+m_T^2+\frac{\Lambda_d^2 v_d^2+\Lambda_u^2 v_u^2}{4}\\
\end{pmatrix}
\,, \notag\\
\mathcal{M}_{21}&= 
\begin{pmatrix}
 v_d ( \sqrt{2}\lambda_d \mud{+} -g_1 M_B^D )\; & \;
v_u (\sqrt{2} \lambda_u\muu{-} +g_1 M_B^D) \\
v_d ( \Lambda_d \mud{+} + g_2 M_W^D) \;& - 
 v_u (\Lambda_u  \muu{-} + g_2 M_W^D) \\
\end{pmatrix}
\;.\notag
\end{align}
Here  $m_A^2=2B_\mu/\sin2\beta$ (with the usual definition of $B_\mu$), we use shorthand notation  $c_\beta\equiv\cos\beta$, $s_\beta\equiv\sin\beta$, $\tan\beta=v_u/v_d$ and 
\begin{align}
\mu_i^{\text{eff,}\pm}
=\mu_i+\frac{\lambda_iv_S}{\sqrt2}
\pm\frac{\Lambda_iv_T}{2}, \;\;
%&\mu_i^{\text{eff,}0}& =\mu_i+\frac{\lambda_iv_S}{\sqrt2}
 i=u,d.
\end{align}
The vacuum expectation values  $v_{d,u,S,T}$ of neutral scalars are assumed real. The MSSM-like 
$\mathcal{M}_{\text{MSSM}}$ and  the singlet-triplet-like $\mathcal{M}_{22}$ submatrices mix  through the  $\mathcal{M}_{21}$ term.

In the analogy to the MSSM, the SM-like Higgs state is considered to be $\phi_u$-dominated while the $\phi_T$ is assumed heavy since a large soft breaking mass term $m_T$ is a natural and sufficient way to suppress the tree-level triplet contribution to the $W$ boson mass and $\rho$ parameter.
This leads to two phenomenologically viable scenarios, with either the SM-like Higgs being the lightest state \cite{Diessner:2014ksa} or the next-to-lightest \cite{Diessner:2015iln}, with a low mass singlet scalar.

To understand how one can realize a scenario with $m_{h_1} \approx 95$ and $m_{h_2} \approx 125$ GeV it is instructive to consider the limit in which not only the triplet decouples \textcolor{red}{(leading to a Higgs sector field content as in the singlet extended 2HDM \cite{Biekotter:2019kde,Aguilar-Saavedra:2023tql,Banik:2023ecr}, albeit with non-trivial relations between couplings)} but also in which the MSSM-like pseudoscalar Higgs mass $M_A$ and the value of $\tan\beta=v_u/v_d$ become large (this last approximation will turn out also to be a necessary condition to fit LEP and LHC signal strengths).
In this limit it is enough to focus on the 2x2 sub-matrix corresponding to the $(\phi_u,\phi_S)$ fields only, which reads
\begin{align}
\mathcal{M}^{\phi}_{u,S}&=
\begin{pmatrix}
m_Z^2 +\Delta m^2_{rad}& v_u \left(\sqrt{2} \lambda_u \muu{-} +g_1 M_B^D\right) \\
v_u \left(\sqrt{2} \lambda_u\muu{-} +g_1 M_B^D\right) \; &
4(M_B^D)^2+m_S^2+\frac{\lambda_u^2 v_u^2}{2} \;\\
\end{pmatrix}
\;,
\label{eq:hu-s-matrix}
\end{align}
where $\Delta m^2_{rad}$ denotes expected large quantum corrections to the mass of the SM-like Higgs.
A setup with a $\sim95$ GeV scalar therefore requires the matrix element (2,2) small compared to the (1,1) one. 
This requires   
\begin{align}
  M_B^D, m_S \lesssim m_Z\; 
\end{align}
and since $v_u\approx v=246$ GeV for large $\tan\beta$, the coupling $\lambda_u$ therefore must be  also very small.
The off-diagonal matrix element $v_u \left(\sqrt{2} \lambda_u \muu{-} +g_1 M_B^D\right)$ must remain small in order not to disturb the properties of the SM-like Higgs state $h_2$.
As we will see in the next section, the dark matter  data will impose additional constraints among these parameters. 

%Before progressing to phenomenologically analysis of proposed benchmark points, 
Let us first comment on original benchmark points BMP4--6 of Ref.~\cite{Diessner:2015iln}.
While the mixing and masses are more-or-less appropriate to fix the LEP excess, none of the points exhibits an enhancement in the $\gamma \gamma$ channel.
Moreover, BMP5 and BMP6 are nowadays excluded by dark matter direct search at LUX-ZEPPELIN \cite{LZ:2022lsv}, while BMP4 is in tension with $p$-value of 0.083 (as computed by \texttt{micrOMEGAs}).
BMP4 and BMP6 are also excluded by stau searches (the most constraining analysis are \cite{ATLAS:2019gti} and \cite{CMS:2020bfa}, respectively) while BMP5 is excluded by jets + MET search \cite{ATLAS:2020syg}.
%and other analysis.

We searched for  new points exhibiting the 95/125 GeV mass pattern and fitting the LEP and CMS excesses (while remaining allowed by experimental constraints) using differential evolution algorithms from \SciPy \cite{2020SciPy-NMeth}.
Mass spectra and Higgs decays where computed using \FS \cite{Athron:2014yba,Athron:2017fvs,Athron:2021kve,Denner:2016kdg,Sjodahl:2012nk,Allanach:2001kg,Allanach:2013kza,Staub:2009bi,Staub:2010jh,Staub:2012pb,Staub:2013tta}, which we interfaced with \SciPy through its \SLHA output \cite{Skands:2003cj,Allanach:2008qq} using \PySLHA\cite{Buckley:2013jua}.

In Tab.~\ref{tab:BMP} we show two representative benchmark points, with decay patterns of the two lightest CP-even Higgses listed in Tab.~\ref{tab:br}.
These benchmarks are characterised by\footnote{\FS inputs as well as obtained spectrum files for the above points are attached to the \texttt{arXiv} version of this work.}

%However, the off-diagonal matrix element contains a (R-)Higgsino mass parameter $\mu_u$, which cannot be small.
%As we will see in the next section, dark matter direct detection put a lower limit depending on the first generation squark masses on $\mu_u \gtrsim 550$ GeV.

\begin{itemize}
 \item mostly a singlet-like Higgs state $h_1$ with the mass of 95.4 GeV and a SM-like state $h_2$ with $m_{h_2} \approx 125.25~\text{GeV}$,
 \item where $h_1$ fits the LEP excess with (approximating production via partial widths)
\begin{align}
\label{eq:muzbb}
\muzbb \approx \frac{\Gamma(h_1 \to ZZ) \text{BR}(h_1\to b\bar{b})}{\Gamma(h^\text{SM}_{95.4} \to ZZ) \text{BR}(h^\text{SM}_{95.4}\to b\bar{b})} = 0.117,
\end{align}
\item and where $h_1$ is produced at the LHC with
\begin{align}
\label{eq:muAA}
\muaa \approx \frac{\Gamma(h_1\to gg) \text{BR}(h_1 \to \gamma \gamma)}{\Gamma(h^\text{SM}_{95.4} \to gg) \text{BR}(h^\text{SM}_{95.4} \to \gamma \gamma)} = 0.24,
\end{align} 
\item due to small bino-singlino mass parameter $M_B^D$  both benchmarks feature a light neutralino, which can serve as a dark matter candidate,
\item however BMP7 and BMP8 differ in how the relic density of DM is achieved, see next section.
\end{itemize}

Since colour and electrically charged particles are (apart for right handed staus in BMP8) fairly heavy ($\gtrsim 0.5$ TeV), the $\muaa$ is not enhanced relative to $\muzbb$ via contribution from new particles in the loop.
Desired values of $\muzbb$ and $\muaa$ are rather achieved almost exclusively via Higgs mixing.
The ratios of $\Gamma(h_1 \to ZZ)/\Gamma(h^\text{SM}_{95.4} \to ZZ)$, $\Gamma(h_1\to gg)/\Gamma(h^\text{SM}_{95.4} \to gg)$,  and $\Gamma(h_1\to \gamma \gamma)/\Gamma(h^\text{SM}_{95.4} \to \gamma \gamma)$ are around 0.13 and have a simple scaling with the $S-H_u$ mixing matrix element.
This makes them fully correlated.
The non-universality, needed to  explain the difference between $\muzbb$ and $\muaa$ comes from ratios of branching ratios in Eqs.\ \ref{eq:muzbb} and \ref{eq:muAA}.
Since the total width of scalar states between 90 -- 130 GeV is dominated by the decay to $b\bar{b}$, the $\text{BR}(h_1\to b\bar{b})$ is fairly insensitive to the mixing effects which cancel between  the nominator and denominator in the branching ratio, giving $\text{BR}(h_1\to b\bar{b}) \approx 0.9\cdot\text{BR}(h^\text{SM}_{95.4}\to b\bar{b})$ when at the same time $\Gamma(h_1\to b\bar{b}) \approx 0.074 \cdot \Gamma(h^\text{SM}_{95.4}\to b\bar{b})$ (note that there is a difference between suppression for the width generated via the $H_d$ and $H_u$ admixtures).
This is no longer the case for the $\text{BR}(h_1 \to \gamma \gamma)$ which gets enhanced in relation to
$\text{BR}(h^\text{SM}_{95.4} \to \gamma \gamma)$ due to a weaker suppression of $\Gamma(h_1 \to \gamma \gamma)/\Gamma(h^\text{SM}_{95.4} \to \gamma \gamma)$ in relation to total width of $h_1$.
This setup is similar to the one described in \cite{Cao:2023gkc} in the NMSSM. However, the required pattern of masses and mixings is far more difficult to realise in the MRSSM since its constraining nature connects different sectors of the model which means that experimental limits from the  dark matter and general collider phenomenology have impact on the Higgs sector.

\begin{figure}
 \centering
 \begin{subfigure}
 {0.49\textwidth}
 \includegraphics[width=\textwidth]{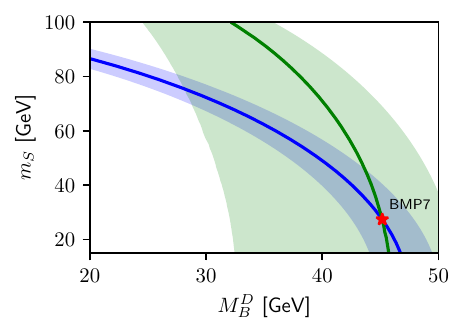}
 \caption{}
 \label{fig:mh_M1_mS}
 \end{subfigure}
 \begin{subfigure}
 {0.49\textwidth}
 \includegraphics[width=\textwidth]{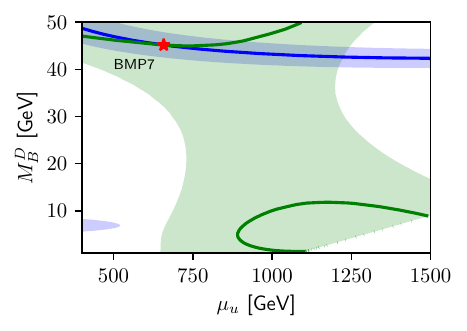}
 \caption{}
 \label{fig:mh_muU_M!}
 \end{subfigure}
 \begin{subfigure}
 {0.49\textwidth}
 \includegraphics[width=\textwidth]{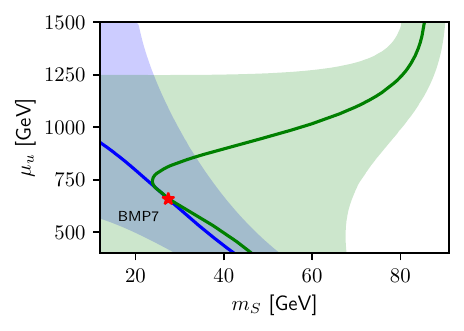}
 \caption{}
 \label{fig:mh_mS_muU}
 \end{subfigure}
  \begin{subfigure}
 {0.49\textwidth}
 \includegraphics[width=\textwidth]{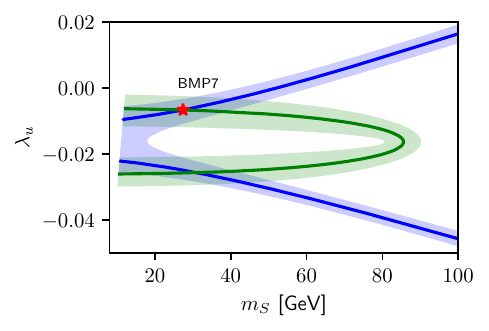}
 \caption{}
 \label{fig:mh_mS_lamSU}
 \end{subfigure}
  \begin{subfigure}
 {0.49\textwidth}
 \includegraphics[width=\textwidth]{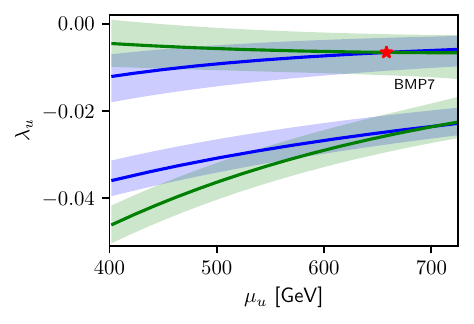}
 \caption{}
 \label{fig:mh_lamSU_muU}
 \end{subfigure}
 \caption{
 Dependence of Higgs masses on parameters controlling the singlet-doublet mixing.
 We mark combinations of parameters giving $m_{h_1} = 95.4$ GeV (blue), $m_{h_2} = 125.25$ GeV (green) around BMP7, with bands marking $\pm 3$ GeV regions.
 }
 \label{fig:masses}
\end{figure}
\begin{figure}
 \centering
 \begin{subfigure}
 {0.49\textwidth}
 \includegraphics[width=\textwidth]{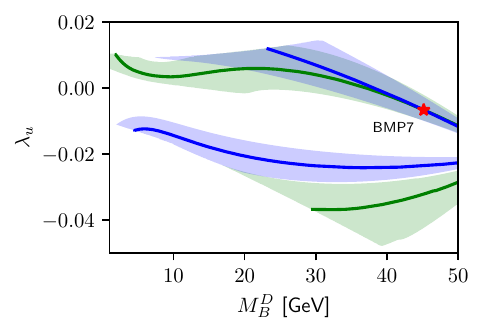}
 \caption{}
 \label{fig:excess_M1_lamSU}
 \end{subfigure}
  \begin{subfigure}
 {0.49\textwidth}
 \includegraphics[width=\textwidth]{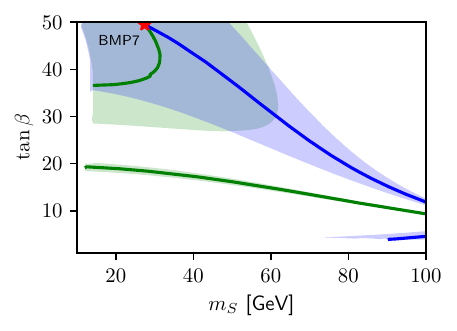}
 \caption{}
 \label{fig:excess_mS_tanb}
 \end{subfigure}
   \begin{subfigure}
 {0.49\textwidth}
 \includegraphics[width=\textwidth]{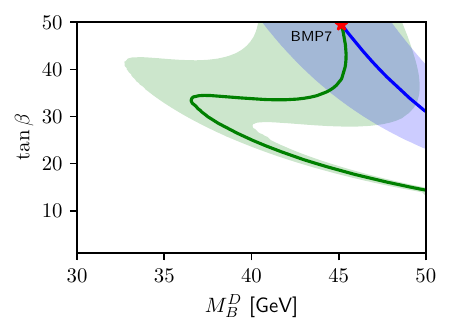}
 \caption{}
 \label{fig:excess_M1_tanb}
 \end{subfigure}
    \begin{subfigure}
 {0.49\textwidth}
 \includegraphics[width=\textwidth]{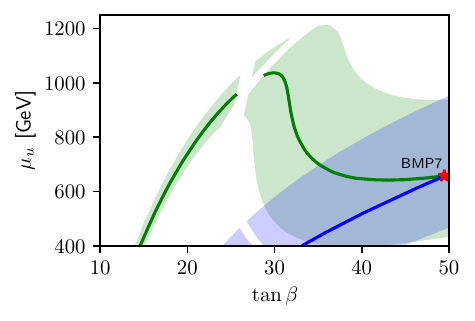}
 \caption{}
 \label{fig:excess_tanb_muU}
 \end{subfigure}
 \caption{
 Dependence of signal strengths on parameters controlling the singlet-doublet mixing.
 Contours mark combinations of parameters giving $\mu_{Zb\bar{b}} = 0.117 \pm 0.057$ (blue), $\mu_{\gamma \gamma} = 0.24^{+0.09}_{-0.08}$ (green) around BMP7, with bands marking $1\sigma$ deviations.
 }
 \label{fig:strengts}
\end{figure}

To understand how the  desired pattern of partial widths  described above is achieved for parameter points given in Tab.~\ref{tab:BMP} we make a set of $2d$ scans around BMP7.
We focus here on masses of two lightest CP-even Higgses (Fig.~\ref{fig:masses}) and LEP and LHC signal strengths (Fig.~\ref{fig:strengts}), relegating the analysis of experimental limits, like dark matter or the consistency of the Higgs sector with experimental data, to Sec. \ref{sec:experimental_constraints}.
In our setup, where the lightest Higgs is mainly a singlet (with mass given approximately by $\sqrt{m_S^2 +4 (M_B^D)^2}$ as show in Fig.~\ref{fig:mh_M1_mS}) and $\mu_u \gtrsim 0.5$ TeV to avoid direct searches of SUSY particles at the LHC (see discussion in Sec. \ref{sec:LHC}), $\lambda_u$ is constrained to a very small range of $\lambda_u \in [-0.05, -0.0025]$ (see Figs.~\ref{fig:mh_mS_lamSU} and \ref{fig:mh_lamSU_muU}) and is strongly correlated with $\mu_u$ (Fig. \ref{fig:mh_lamSU_muU}), as it appears as a $\lambda_u \mu_u^{\text{eff},-}$ product in the $(2,1)$ entry of the mass matrix in Eq.~\ref{eq:hu-s-matrix}.
The light CP-even state with mass of 95 GeV could be  also achieved for a larger value of $|\lambda_u|$ when larger mixing with the SM-like state compensates an increased value of $\sqrt{m_S^2 +4 (M_B^D)^2}$,  but such points would then be excluded by the Higgs sector analysis in Sec.~\ref{sec:HT}.
Therefore, for a given range of $\lambda_u$,  the parameter $\mu_u$ influences non-trivially mainly the SM-like Higgs mass via one-loop corrections, where $\mu_u$ can appear without the $\lambda_u$ parameter.
The situation is somewhat similar for LHC and LEP signal strengths.
With a small allowed region of $\lambda_u$ (Fig.~\ref{fig:excess_M1_lamSU}) and rather simple functional dependence of $\mu_{Zb\bar{b}}$ (Figs.  \ref{fig:excess_mS_tanb}, \ref{fig:excess_M1_tanb} and \ref{fig:excess_tanb_muU}), 
the non-trivial dependence appears mainly for $\mu_{\gamma \gamma}$ as it is loop induced.
For other combinations of parameters appearing in Eq.~\ref{eq:hu-s-matrix} but not shown in Fig.~\ref{fig:masses} or Fig.~\ref{fig:strengts} masses or signal strength, respectively, are mostly correlated not leading to any non-trivial relations. 

\begin{table}[t]
\begin{center}
\begin{tabular}{l|l|l}
& BMP7 & BMP8 \\
\midrule
\midrule
$\tan\beta$ & $49.5$ & $49.8$ \\
$B_\mu$     & $176^2$ & $142^2$ \\
$\lambda_d$, $\lambda_u$ & $-0.193, -0.00658 $ & $0.161,-0.0135 $ \\
$\Lambda_d$, $\Lambda_u$ & $1.49,-1.03$ & $1.49, -0.722$ \\
%\midrule
$M_B^D$ & $45.2$ & $42.1$ \\
$m_S^2$ & $27.4^2$ & $54.1^2$ \\
$m_{R_u}^2$, $m_{R_d}^2$& $1292^2,522^2$ & $1033^2$,$788^2$ \\
$\mu_d$, $\mu_u$ & $1536,658$ & $1500,1282$ \\
$M_W^D$ & $1458$ & $1490$ \\
$M_O^D$ & \multicolumn{2}{c}{$3000$} \\
$m_T^2$, $m_O^2$ & \multicolumn{2}{c}{$3000^2,1500^2$} \\
%\midrule
$m_{Q;1,2}^2$, $m_{Q;3}^2$ & $3803^2,3900^2$ & $1465^2, 3477^2$\\
$m_{D;1,2}^2$, $m_{D;3}^2$ & $3148^2,3728^2$ & $1456^2, 1990^2$\\
$m_{U;1,2}^2$, $m_{U;3}^2$ & $1271^2,2452^2$ & $3285^2, 3967^2$\\
$m_{L;1,2}^2$, $m_{E;1,2}^2$&$1000^2$, $1000^2$ & $1680^2$, $1022^2$\\
$m_{L;3,3}^2$, $m_{E;3,3}^2$ & $1000^2$, $1000^2$& $803^2$, $185^2$\\
\midrule
$m_{H_d}$ & $-1884^2$ & $-1711^2$\\
$m_{H_u}$ & $-1063^2$ & $-1534^2$\\
$v_S$ & $-3087$ & $2004$ \\
$v_T$ & $0.35$ & $0.0142 $\\
\midrule
\midrule
$m_{h_1}$  & 95.4 & 95.4 \\
$m_{h_2}$  & 125.25 & 124.72 \\
$m_{W^\pm}$ & 80.375 & 80.371 \\
$m_{\chi_1}$  & 44.98 & 42.65 \\
$m_{\tilde{\tau}_R}$ & 1000 & 124.7 \\
$\rho^\pm_1$ & 717 &  1310 \\
$m_{a}$  & 24.85 & 54.20 \\
\bottomrule
\end{tabular}
\end{center}
\caption{
Benchmark points for the scenario discussed here: input parameters, parameters determined via tadpole equations and selected predicted, phenomenologically relevant, pole masses. Dimensionfull parameters are given in GeV or GeV${}^2$, as appropriate.
Input values %,  rounded to 3 or 4 significant digits 
are listed in the upper part of the table, while derived masses of some  light physical states are in the lower part.
\FS input/output files (including more significant digits of input parameters) are attached to the \texttt{arXiv} version of this work.
}
\label{tab:BMP}
\end{table}
\begin{table}
\begin{center}
\begin{tabular}{c|cc|cc}
 & \multicolumn{2}{c}{BMP7} & \multicolumn{2}{c}{BMP8}\\
 \hline
decay channel  & $h_1$ & $h_2$ & $h_1$ & $h_2$\\
 \hline
 $b \bar{b}$ & 72.0 & 56.8 & 71.9 & 58.2 \\
 $W^+ W^-$ & 0.874 & 22.0 & 0.877 & 20.7 \\
 $g g$ & 12.5 & 8.97 & 12.5 & 8.86 \\
 $\tau^+ \tau^-$ & 7.28 & 6.08 & 7.27 & 6.22 \\
 $c \bar{c}$ & 6.93 & 2.93 & 6.96 & 2.92 \\
 $ZZ$ & 0.125 & 2.82 & 0.125 & 2.64 \\
 $\gamma \gamma$ & 0.253 & 0.252 & 0.253 & 0.250 \\
 $\gamma Z$ & $9.84\cdot10^{-4}$ & 0.159 & $9.91 \cdot 10^{-6}$ & 0.152 \\
 $\chi_1 \bar{\chi}_1$ & $7.31\cdot 10^{-4}$ & $2.88 \cdot 10^{-3}$ & $5.14 \cdot 10^{-5}$ & $9.82 \cdot 10^{-3}$\\
 $a a$ & $9.33 \cdot 10^{-4}$ & $3.03 \cdot 10^{-4}$ & 0 & $1.99 \cdot 10^{-3}$ \\
 \hline
 total width [MeV] & 0.173 & 3.41 & 0.172 & 3.41
\end{tabular}
\end{center}
\caption{Widths and branching ratios (in \%) of the 95 and 125 GeV Higgs states for benchmark points BMP7 and 8 (see Tab.~\ref{tab:BMP}).
%Some channels with BR larger than $\gamma Z$ but of little phenomenological importance were omitted.
More complete spectrum files are attached to the \texttt{arXiv} version of this work.
}
\label{tab:br}
\end{table}
We should mention that neither BMP7 nor BMP8   fit the small  $\mu_{\tau \tau}$ excess at 95 GeV.  
This excess cannot be achieved via the mixing while fitting at the same time $\muzbb$ as in the THDM type II, where $h_1$ couplings to $b\bar{b}$ and $\tau \bar{\tau}$ are fully correlated (from the point of view of the pure Higgs sector, our setup is analogues to the type II version of S2HDM from Ref.~\cite{Biekotter:2023jld}).
This can be revisited in the future when \FS will be able to compute loop corrections to the $\tau \bar{\tau}$ decay, especially in a scenario with light staus, although expecting such large corrections seems unreasonable.

\section{Confronting with experimental observations}
\label{sec:experimental_constraints}

In this section we discuss the phenomenological properties of benchmark points proposed in Sec.~\ref{sec:mrssm}.
In particular, we asses their viability in light of constraints coming from Higgs sector measurements, dark matter properties and general collider phenomenology.
The properties of predicted Higgs bosons are checked against experimental data using \HT v1.1.4 \cite{Bahl:2022igd}, whereas dark matter properties and LHC constraints are checked using the combination of \MO v6.0 \cite{Barducci:2016pcb,Belanger:2018ccd,Alguero:2023zol} and \texttt{SModelS} v2.3.3 \cite{Ambrogi:2017neo,Ambrogi:2018ujg,Dutta:2018ioj,Altakach:2023tsd}, based on \SA generated \texttt{CalcHEP} model file \cite{Staub:2009bi,Belyaev:2012qa}.
We show a selection of 2-dimensional scans around the BMPs to explain which values of parameters are necessary to avoid current experimental limits.
Finally, in Sec.~\ref{sec:LHC} we comment briefly on prospects for discovering of the MRSSM at future collider experiments.

\subsection{Higgs sector}
\label{sec:HT}

\begin{table}[bt]
\begin{center}
\begin{tabular}{lcc}
& BMP7 & BMP8 \\
\hline
\texttt{HiggsSignals} $p$-value  & 0.120 & 0.0586 \\
\texttt{HiggsBounds} $h_1$ \texttt{obsRatio}  & 0.40 & 0.40 \\
\hline 
$\Omega h^2$ &  0.121 & 0.121 \\
direct detection $p$-value & 0.5 & 0.5\\
\hline 
\texttt{SModelS} $r_{95}$ & 0.86 & 0.57
\end{tabular}
\end{center}
\caption{Summary of experimental constraints passed by BMPs.}
\label{tab:ht}
\end{table}

\begin{figure}
 \centering
 \begin{subfigure}
 {0.49\textwidth}
 \includegraphics[width=\textwidth]{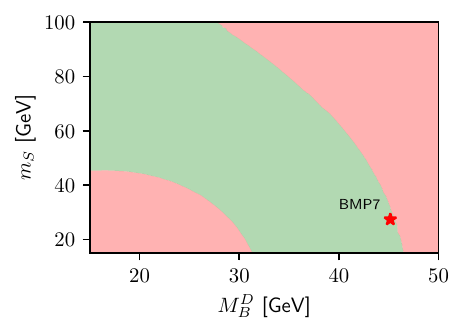}
 \caption{}
 \label{fig:hs_bmp1_M1_mS}
 \end{subfigure}
 \begin{subfigure}
 {0.49\textwidth}
 \includegraphics[width=\textwidth]{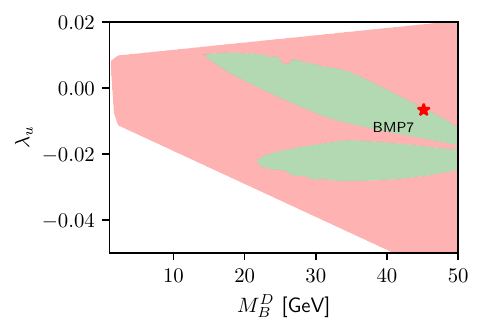}
 \caption{}
 \label{fig:hs_bmp1_M1_lamU}
 \end{subfigure}
 \caption{
    Parameter regions around BMP7 allowed (green) and excluded (red)  by SM-like Higgs data at 95\% C.L. as reported by \HS.
    White regions is where no spectrum could be generated.
 }
\label{fig:hs}
\end{figure}

\begin{figure}
 \centering
 \begin{subfigure}
 {0.49\textwidth}
 \includegraphics[width=\textwidth]{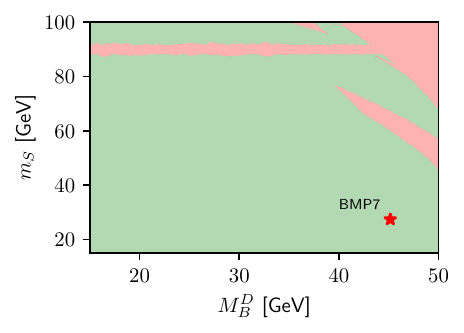}
 \caption{}
 \label{fig:hb_bmp1_M1_mS}
 \end{subfigure}
 \begin{subfigure}
 {0.49\textwidth}
 \includegraphics[width=\textwidth]{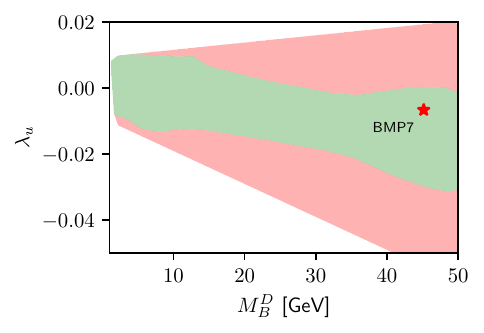}
 \caption{}
 \label{fig:hb_bmp1_M1_lamU}
 \end{subfigure}
 \caption{
    Parameter regions around BMP7 allowed (green) and excluded (red) by searches of non SM-like Higgses at 95\% C.L.  as reported by \HB.
    White regions is where no spectrum could be generated.
 }
\label{fig:hb}
\end{figure}

The Higgs sector of the model was checked against experimental data using \FS interface to \HT.\footnote{The release of \FS interface to \HT and \LL is in preparation. A preliminary version can be obtained from \href{https://github.com/FlexibleSUSY/FlexibleSUSY/tree/feature-HiggsTools-interface}{https://github.com/FlexibleSUSY/FlexibleSUSY/tree/feature-HiggsTools-interface}.}
The validity of any parameter point is assessed by comparing its $\chi^2$ with the minimal $\chi^2$ in the SM. In the SM we find the minimal value  $\chi^2_{\text{SM}} = 151.55$ at the Higgs boson mass of 125.25 GeV (all of the numbers use \texttt{HiggsSignals} database v1.1 and assume 3\% uncertainty on the Higgs mass prediction in the BSM model).
% see http://www.reid.ai/2012/09/chi-squared-distribution-table-with.html
% for 2 d.o.f. 3σ corresponeds to delta chi^2 < 11.83
% actually it's 5.991464547107979
The  allowed region in the parameter space at 95\% C.L.\ corresponds to $\Delta \chi^2 \equiv \chi^2 - \chi^2_{\text{SM}} \lesssim 5.99$.\footnote{This criterion for calculating the allowed region is recommended by Ref.~\cite{Biekotter:2023oen}}
Tab.~\ref{tab:ht} shows the $p$-values for both BMPs computed from this $\Delta \chi^2$ for 2-degrees of freedom.\footnote{BMPs were also checked for SM-like Higgs bosons using \LL \cite{Bernon:2015hsa,Kraml:2019sis,Bertrand:2020lyb} v2.1 with database 22.01-beta.1 obtaining $p$-values of 0.159 (BMP7) and 0.0846 (BMP8).}
Both points are within $2\sigma$, though more than $1\sigma$ away,  so if the LHC anomalies persist while the measurements of SM-like Higgs boson improve, this solution might become disfavoured.
The points also predict a roughly SM-like $h_2 Z \gamma$ coupling, which is of no consequence now \cite{ATLAS:2023yqk} but might become important once the precision of this measurement increases and statistically significant deviation is observed.

In Fig.\ \ref{fig:hs} we show example regions allowed by measurements of SM-like Higgs properties for selected parameters relevant to properties of the 95 GeV scalar.
While properties of the second-to-lightest Higgs boson are determined mostly by other parameters, which were chosen to make it mostly SM-like, the mixing with the singlet can destabilize them.
Since some mixing is needed to give the 95 GeV state its couplings to SM particles, there is obviously also an upper limit on it.
This is seen for example in Fig.\ \ref{fig:hs_bmp1_M1_lamU} where the increase of $M_B^D$, which appears in the (2,1) element of the mass matrix in Eq.~\ref{eq:hu-s-matrix} as a combination $\sqrt{2} \lambda_u\muu{-} +g_1 M_B^D$, has to be compensated by a larger value of $|\lambda_u|$.

Properties of the non-SM Higgs with the mass of 95 GeV were checked using a \HB component of \HT (database version 1.4).
This state is mostly constrained by the wanted 95 GeV LEP anomaly \cite{LEPWorkingGroupforHiggsbosonsearches:2003ing}, with the ratio of production cross sections to the 95\% limit (referred by \texttt{HiggsBounds} as \texttt{obsRatio}) being 0.4 (Tab.~\ref{tab:ht}).
In Fig.\ \ref{fig:hb} we show example 2$d$ regions around BMP7 allowed by \HB.
When the admixture of SM-like Higgs in the lightest state becomes too large the points start to get excluded by searches of SM or non-SM like Higgses (e.g. \cite{CMS:2018cyk} or \cite{CMS:2018amk}) as the lightest state couples more-and-more strongly to SM particles and the second-to-lightest Higgs starts to deviate from the SM.

Since the  remaining charged and CP-even neutral scalar states are fairly heavy, they evade any experimental bounds.
The only Higgs-like states which are light are the  lightest CP-odd scalars with masses $m_{a} = 24.85$ (BMP7) and 54.20 (BMP8) GeV.
Those states are fairly long lived with partial widths $10^{-12}-10^{-11}$ GeV and couple almost exclusively to $\gamma \gamma$.
Thus they cannot  easily be produced at $pp$ and $e^+e^-$ colliders   via the  $\gamma \gamma$ channel, since 
the production  cross-section is extremely small and therefore they easily evade all experimental bounds.

\subsection{Dark matter sector}

As explained in Sec.~\ref{sec:mrssm}, the light-singlet setup with the bino-singlino mass parameter $m^D_B \sim 60$ GeV in the MRSSM inevitable leads to a light (lighter than around 60 GeV) dark matter candidate (LSP), making the DM relic density and direct detection limits  important constraints.
The two BMPs in Tab.~\ref{tab:BMP} were selected such as to represent different  ways  how the correct relic density is achieved. For BMP7 the LSP  correct relic density is achieved by  the $s$-channel resonant  LSP pair annihilation into $Z$ bosons,  while  in BMP8 stau is light enough to   annihilate LSP pairs  via $t$-channel $\tilde{\tau}_R$ exchange into tau leptons. 
The  direct detection expectations crucially depend on the  relation between  the $\mu_u$ parameter and the  first generation squark masses.

We compute dark matter observables with \MO using \SLHA output generated by \FS (\FS output for the MRSSM is fully compatible with \SA generated \texttt{CalcHEP}/\MO model with the caveat that one has to call the output \SLHA file \texttt{SPheno.spc.MRSSM}).

\subsubsection{Relic density}

\begin{figure}
 \centering
 \begin{subfigure}
 {0.49\textwidth}
 \includegraphics[width=\textwidth]{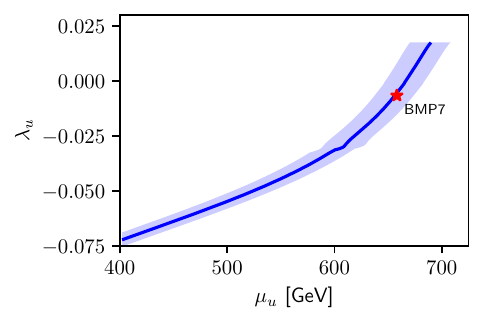}
 \caption{}
 \label{fig:relic_muu_lamu}
 \end{subfigure}
 \begin{subfigure}
 {0.48\textwidth}
 \includegraphics[width=\textwidth]{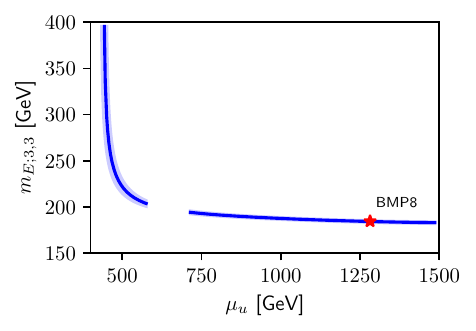}
 \caption{}
 \label{fig:relic_M1_mSe}
 \end{subfigure}
 \begin{subfigure}
 {0.49\textwidth}
 \includegraphics[width=\textwidth]{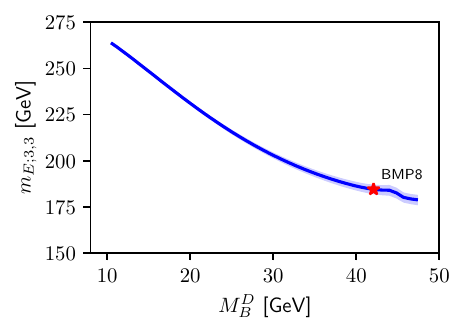}
 \caption{}
 \label{fig:relic_muu_mSe}
 \end{subfigure}
 \caption{
 Contours showing combinations of parameters giving $\Omega h^2 = 0.12$ around BMP7 (\ref{fig:relic_muu_lamu}) and BMP8 (\ref{fig:relic_M1_mSe}, \ref{fig:relic_muu_mSe}), with bands marking parameters giving relic density up to $\pm 10\%$ away from this target. 
 Without light staus, the annihilation occurs solely via \mbox{(R-)H}iggsino admixture, making $\mu^{\text{eff}}_u \approx \mu_u + \tfrac{1}{\sqrt{2}} \lambda_u v_S$ a crucial parameter controlling the relic density (\ref{fig:relic_muu_lamu}).
 This also fixes $m_\chi \approx m_Z/2$.
 Light staus open the range of DM masses allowed by relic density constraint (\ref{fig:relic_M1_mSe}) and make it insensitive to the value of $\mu_u$ (\ref{fig:relic_muu_mSe}).
 This parameter is however still important for  avoiding direct detection constraints.
 }
\end{figure}

As we said, the correct relic density in the setup with $m_\chi \lesssim m_Z$ can be achieved in two ways: $s$-channel annihilation via resonant $Z$ (BMP7) or annihilation via $t$-channel $\tilde{\tau}_R$ (BMP8):

BMP7: in this scenario with heavy staus, the correct relic density can only be achieved if $m_{\chi_1} \sim m_Z/2$.
In the MRSSM the $Z$ boson couples to neutralinos only via 
their (R-)Higgsino admixture.
This is controlled by $\tan \beta$ and the relevant $\mu_i^{\text{eff,}\pm}$ parameter (where the dependence on $\tan \beta$ is negligible so long as $\tan \beta$ is not $\sim 1$).
With $M_B^D \lesssim m_Z$ and the size of the off-diagonal mixing determined by the electroweak scale this puts a strong bounds on the size of $\mu_u^{\text{eff,}-}$.
The smallness of $\mu_u^{\text{eff,}-} \approx \mu_i + \tfrac{1}{\sqrt{2}}\lambda_i v_S$ can be achieved by making both elements  small or by relaying on mutual cancellation.
Since $v_S$ is large  in the setup with light singlet, it forces $\lambda_u$ to be relatively small if no artificial cancellation between $\mu_u$ and $\lambda_u v_S$ is enforced.
The dependence of relic density on $\mu_u$ and $\lambda_s$ is shown in Fig.~\ref{fig:relic_muu_lamu}.
This scenario fixes $\mu_u$ to be in a very restrictive range of 600 -- 700 GeV.

BMP8: 
In this scenario  $m_{\tilde{\tau}_R} \sim 100 - 200$ GeV and the majority (98\% in case of BMP8) DM annihilation happens through $\chi_1 \bar{\chi}_1 \to \tau^+ \tau^-$.
This opens the range of $M_B^D$ allowed by relic density as shown in Fig.~\ref{fig:relic_M1_mSe}.
$\mu_u$ is no longer as constrained since the annihilation happens also via the Bino part of $\chi_1$ (Fig.~\ref{fig:relic_muu_mSe}).
Higgs constraints still force $M_B^D \sim m_Z/2$ though.

\begin{figure}
 \centering
 \begin{subfigure}
 {0.49\textwidth}
 \includegraphics[width=\textwidth]{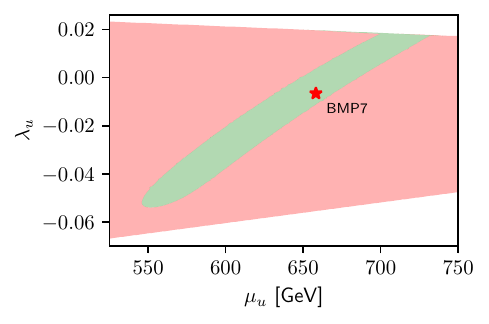}
 \caption{}
 \label{fig:DD_BMP1_muU_lamSU}
 \end{subfigure}
 \begin{subfigure}
 {0.47\textwidth}
 \includegraphics[width=\textwidth]{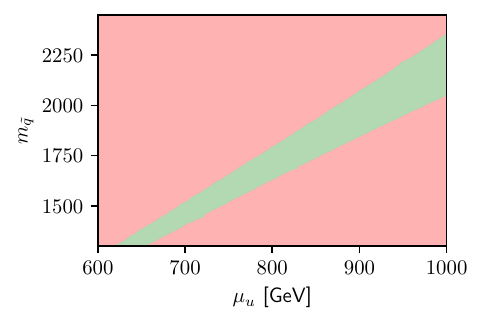}
 \caption{}
 \label{fig:BMP1_DM_DD_muU_msq}
 \end{subfigure}
 \caption{
Parameter regions around BMP7 allowed (green) and excluded (red) by dark matter direct detection experiments at 95\% C.L.
White regions is where no spectra could be generated.
     Since in Fig. \ref{fig:BMP1_DM_DD_muU_msq} all first and second generation soft squark masses were set equal, $m_{Q;1,2} = m_{U;1,2} = m_{D;1,2} \equiv m_{\tilde{q}}$,  to facilitate scanning, 
 none of the points corresponds directly to BMP7 or BMP8. 
 }
\end{figure}

\subsubsection{Direct detection}
\label{sec:DM}
If dark matter annihilation happens via the $Z$ exchange, the same diagram contributes to the scattering of dark matter in  direct detection experiments, making it sensitive to the same parameters as relic density.
In Fig.\ \ref{fig:DD_BMP1_muU_lamSU} we show the region allowed by direct detection in the $\mu_u - \lambda_u$ plane.
To avoid  current bounds one has to invoke destructive interference with first generation squarks leading to a strong correlation between $\mu_u$ and $m_{\tilde{q}}$ as seen in Fig.\ \ref{fig:BMP1_DM_DD_muU_msq}.
Even in the case when Higgsino component is not relevant for the relic density, it still plays a role in direct detection, also for $\mu_u > 1$ TeV leading to a $\mu_u$ -- $m_{\tilde{q}}$ correlation also in the case of BMP8.

\subsection{Direct collider constraints}
\label{sec:LHC}

In Tab.\ \ref{tab:BMP} we list masses of selected BSM particles.
The two light states in our setup which are in danger of being excluded by direct searches at the LHC are the lightest $\rho^+$ chargino, right handed staus and squarks.
The remaining, collider relevant states (including MRSSM specific ones like Dirac gluinos or color octet scalars) were chosen to be heavy as their masses do not influence observables we are interested in  this work.

Collider limits from direct production of BSM particles at the LHC were checked with \texttt{SModelS} (analysis database v2.3.0, official) using output provided by \texttt{micrOMEGAs}.\footnote{We use \texttt{SModelS} settings closely following the default settings provided by \texttt{micrOMEGAs}. In particular, we do not combine signal regions (SRs) for a given analysis (see Ref.~\cite{Alguero:2020grj} for a description of this feature), reporting only the strongest exclusions. Turning this option on increases runtime by few orders of magnitude, making it problematic for scans. Combining signal regions would put BMP7 slightly above the 95\% exclusion (with observed $r_{95}=1.07$ as apposed to $0.86$, cf. Tab.~\ref{tab:ht}). Such combination increases mostly only the lower limit on $\mu_u$, and as such has no effect on BMP8.}

\begin{figure}
 \centering
  \begin{subfigure}
 {0.465\textwidth}
 \includegraphics[width=\textwidth]{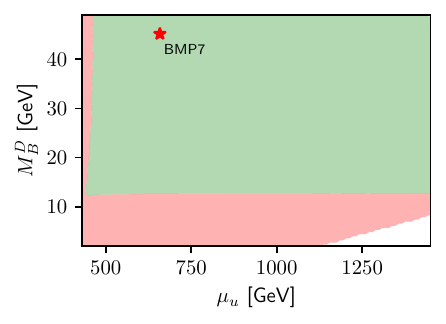}
 \caption{}
 \label{fig:smodels_BMP7_muU_M1}
 \end{subfigure}
 \begin{subfigure}
 {0.49\textwidth}
 \includegraphics[width=\textwidth]{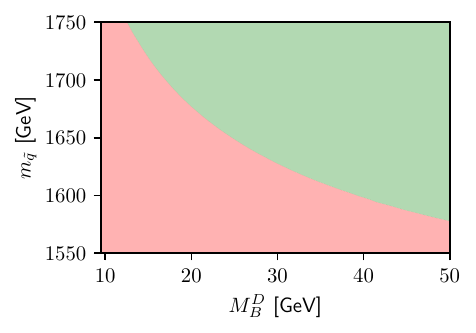}
 \caption{}
 \label{fig:smodels_BMP7_M1_msq}
 \end{subfigure}
  \begin{subfigure}
 {0.49\textwidth}
 \includegraphics[width=\textwidth]{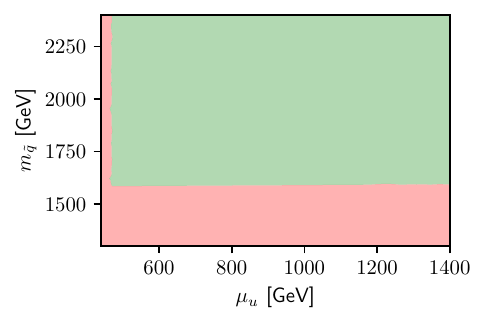}
 \caption{}
 \label{fig:smodels_BMP7_muU_msq}
 \end{subfigure}
 \caption{Regions allowed (green) and excluded (red) by direct searches of BSM particles at 95\% C.L. as given by \texttt{SModelS}.
 White regions is where no spectrum could be generated.
 Since in Fig. \ref{fig:smodels_BMP7_M1_msq} and \ref{fig:smodels_BMP7_muU_msq} all first and second generation soft squark masses were set equal, $m_{Q;1,2} = m_{U;1,2} = m_{D;1,2} \equiv m_{\tilde{q}}$, to facilitate scanning, 
 none of the points corresponds directly to BMP7 or BMP8.
 }
\end{figure}

As shown in Fig.\ \ref{fig:smodels_BMP7_muU_M1}, there is a lower limit on $\mu_u$ of around 500 GeV, making it low enough to allow the values of $\mu_u$ necessary for a correct DM relic density.
In Sec.\ \ref{sec:DM} we have determined the first generation squark masses needed in the case of $\mu_u \sim 650$ GeV.
Such masses are currently slightly above the current experimental limit, as shown in Fig. \ref{fig:smodels_BMP7_M1_msq}.
We note that the limits in Fig.\ \ref{fig:smodels_BMP7_M1_msq} are stronger than actually needed, as we assumed there that all masses of first 2 generation squarks ($m_Q$, $m_D$, $m_U$) are equal.
In reality, as can be seen in Tab.~\ref{tab:BMP}, we require that only some of them are small.

Finally, as is important in the case of BMP8, right handed staus with 100\% branching ratio $\tilde{\tau}_R \to \tau \chi_0$ are excluded at 95\% CL for masses $< 89.8$ \cite{OPAL:2003nhx}.
At LHC they are mostly unconstrained \cite{CMS:2022syk} (which is also the result reported by \texttt{SModelS}).

In the future, light scalars will be searched for not only at the LHC but also at the planned $e^+e^-$ colliders.
Considered $e^+e^-$ machines like CLIC, ILC, FCCee, CEPC and C$^3$ would allow to precisely measure properties of the 95 GeV state in the "Higgs-strahlung" process \cite{Drechsel:2018mgd,Zarnecki:2024xdy,Wang:2020lkq,Mekala:2020zys,Mekala:2021uvg}.
The search of light scalar in the "Higgs-strahlung" process was also defined as on of focus topics for the ECFA study on Higgs/Top/EW factories \cite{deBlas:2024bmz}.
Definite disentanglement of the MRSSM from other models featuring a singlet extended Higgs sector would however require a discovery of some of the MRSSM specific particles like R-Higgses.
They would be pair produced at both lepton and hadron colliders decaying into pairs of R-charge 1 SUSY particles, e.g. $R^- \to \chi \chi^-$ or $R^0 \to \tilde{l}_L \tilde{l}^*_R$ \cite{Choi:2010an}.
This would result in a interesting but experimentally very challenging signature with long decay chains and multi-particle final states.
If there are other R-charge 1 particles than neutralinos one would expect to observe their direct production first if kinematically allowed.
Therefore the other option to narrow down the MRSSM is by focusing on its $N=2$ gauge sector by looking for sgluons or the Dirac vs. Majorana nature of neutralinos and gluinos.
Sgluons appear in few BSM extensions and are often studied in the context of effective models (see for example \cite{Darme:2024epi}).
However the distinct feature of the MRSSM is that the lightest sgluon, which decays usually directly to top quarks \cite{Kotlarski:2016zhv}, is a pseudoscalar which could be identified by the top quark decay product distribution.
Meanwhile, the Dirac nature of gauginos could be established by looking at their direct production or by their impact on production of other BSM states (via an intermediate $s$ or $t$ channel Dirac gaugino) \cite{Choi:2008pi,Choi:2010gc,Diessner:2019bwv,Borschensky:2024zdg,Chalons:2018gez}.

\section{Summary and conclusions}
\label{sec:conclussions}

The Minimal R-symmetric Supersymmetric Standard Model is a highly predictive type of supersymmetric model, featuring an unbroken $U_R(1)$ R-symmetry at the electroweak scale.
Its constraining nature alters the phenomenology and makes it  different from popular SUSY models like the MSSM or the NMSSM, due to a presence of Dirac gauginos, no left-right sfermion mixing, colour octet scalars and other distinct features.
In the past it was show that this is reflected in a very distinct Higgs, lepton and quark flavour violation, muon magnetic moment and general collider phenomenologies.

Motivated by emerging hints of a low lying scalar resonance we have demonstrated in this work that the MRSSM can accommodate both of the excesses observed  at around 95 GeV at LEP in $e^+e^- \to Z b \bar{b}$ and LHC in $pp \to \gamma \gamma$.
While the setup of the Higgs sector is at least partially similar to a set of singlet-extended 2HDMs, the non-trivial issue is how to realise the necessary pattern of masses and Higgs mixing within the constrains coming from different sectors of the MRSSM.
This is especially obvious in comparison with non-supersymmetric models, where one has a complete freedom in setting masses and mixing of Higgs bosons independently of other parameters.
In contrast, in the MRSSM  the parameters controlling the  Higgs 
mass matrix are  intrinsically connected to  dark 
matter sector.
Light CP-even Higgs leads to an even lighter dark matter candidate.
Relic density and direct detection experiments determine the 
MRSSM's bino-singlino mass parameter $M_B^D$ to be $\sim$45 GeV and 
fix a relation between up-(R)Higgsino mass parameter $\mu_u$ and  
the first generation squark masses.The  
$\mu_u$ and $M_B^D$ parameters in turn influence the  tree-level 
mixing between the light scalar state and the SM-like Higgs, which 
implies that a simultaneous  successful fit to all these observables 
is a non-trivial exercise.

In this work we have identified a region of parameter space where 
the LEP and LHC excesses and dark matter relic density can be 
simultaneously accommodated while evading current experimental 
constraints from LHC and dark matter direct detection experiments 
and predicting SM-like Higgs boson in agreement with current 
measurements.

Finally, we have used this opportunity to showcase our new interface 
between \FS and \texttt{HiggsTools}, allowing for a seamless 
validation of Higgs sectors of a broad class of user defined 
(supersymmetric and non-supersymmetric) models.

\addsec{Acknowledgements}
We thank Henning Bahl and Sven Heinemeyer for their help regarding \HT as well as Alexander Voigt for his constant work on \FS.

JK was supported by the Norwegian Financial Mechanism for years 2014--2021, under the grant No.~2019/\allowbreak34/\allowbreak H/\allowbreak ST2/\allowbreak00707.
WK was supported by the National Science Centre (Poland) grant SONATA No.~2022/\allowbreak47/\allowbreak D/\allowbreak ST2/\allowbreak03087 and, during early stages of this work, by the German Research Foundation (DFG) under grants number STO 876/2--2 and STO 876/4--2.

JK thanks the CERN Theory Division  for hospitality during the final stage of this work.

The authors are grateful to the Centre for Information Services and High Performance
Computing [Zentrum f\"ur Informationsdienste und Hochleistungsrechnen (ZIH)] TU Dresden for providing its facilities for high throughput calculations.

\appendix

\printbibliography[title={References}]
\addcontentsline{toc}{section}{References}

\end{document}